# DATA-CENTRIC MIXED-VARIABLE BAYESIAN OPTIMIZATION FOR MATERIALS DESIGN

**Akshay Iyer, Yichi Zhang**
Mechanical Engineering
Northwestern University, USA

**Aditya Prasad**
Material Science & Engineering
Rensselaer Polytechnic Institute, USA

**Siyu Tao, Yixing Wang**
Mechanical Engineering
Northwestern University, USA

**Linda Schadler**
Engineering & Mathematical Sciences
The University of Vermont, USA

**L Catherine Brinson**
Mechanical Engineering & Material Science
Duke University, USA

**Wei Chen[1]**
Mechanical Engineering
Northwestern University, USA

## ABSTRACT

*Materials design can be cast as an optimization problem with the goal of achieving desired properties, by varying material composition, microstructure morphology, and processing conditions. Existence of both qualitative and quantitative material design variables leads to disjointed regions in property space, making the search for optimal design challenging. Limited availability of experimental data and the high cost of simulations magnify the challenge. This situation calls for design methodologies that can extract useful information from existing data and guide the search for optimal designs efficiently. To this end, we present a data-centric, mixed-variable Bayesian Optimization framework that integrates data from literature, experiments, and simulations for knowledge discovery and computational materials design. Our framework pivots around the Latent Variable Gaussian Process (LVGP), a novel Gaussian Process technique which projects qualitative variables on a continuous latent space for covariance formulation, as the surrogate model to quantify "lack of data" uncertainty. Expected improvement, an acquisition criterion that balances exploration and exploitation, helps navigate a complex, nonlinear design space to locate the optimum design. The proposed framework is tested through a case study which seeks to concurrently identify the optimal composition and morphology for insulating polymer nanocomposites. We also present an extension of mixed-variable Bayesian Optimization for multiple objectives to identify the Pareto Frontier within tens of iterations. These findings project Bayesian Optimization as a powerful tool for design of engineered material systems.*

## 1. INTRODUCTION

The launch of the Material Genome Initiative (MGI) [1] has revolutionized the way advanced material systems are designed with targeted performance. Moving away from the traditional experiment-based cause-effect approach, MGI strives to elucidate the relationship between Processing-Structure-Property (PSP) [2] domains for material design. It requires development of new methods within each of the three domains and protocols to manage information flow across domains. A holistic design strategy for bi-directional traversal of PSP relations requires us to address certain key issues – cost effective processing techniques, microstructure representation and reconstruction, dimensionality reduction and tractable optimization techniques, to name a few.

The emergence of open-source material databases [3-6] and the increasing popularity of machine learning techniques are accelerating our ability to address some of these challenges using a data-centric approach. NanoMine [5, 6], a nanocomposite material database with in-built data curation, exploration and analysis capabilities, represents this approach in the field of polymer nanocomposites. It collects nanocomposite characteristics reported in the literature and from individual research labs including microstructure, processing conditions, and material properties. An ontology-enabled knowledge graph framework helps NanoMine establish relationships between those characteristics. A collection of module tools for microstructure characterization & reconstruction and simulation software to model bulk nanocomposite material response augments knowledge generated by experimental data. Integrating these different sources of knowledge is critical for materials design. However, generating experimental or simulated data for the vast design space defined by the almost infinite combinations of constituents, microstructure morphology, and processing conditions is impractical. This signifies the need for data-centric methodologies that can effectively interrogate existing data and guide "on-demand" computer simulations and physical experiments to accelerate the search of new high performing materials.

---

[1] Corresponding Author email: weichen@northwestern.edu

A range of optimization-based techniques have been developed to support the design of microstructural material systems. Among the existing methods developed by the author group, Physical Descriptors based [7, 8] and Spectral Density function [9, 10] based microstructure designs are the most commonly used approaches due to their physically meaningful characterization, relative ease of reconstruction and low dimensional representation. Recently, we have shown deep learning based Generative Adversarial Networks (GAN) to be effective tools for low dimensional microstructure representation and design [11]. Convolutional layers in GAN can capture higher order spatial correlations in complex morphologies, but the requirement of several thousand microstructures for training presents a barrier.

As microstructural design involves expensive microstructure simulations to assess material properties, most of the existing methods rely on building surrogate models (global metamodels) for replacing the physics-based simulations in parametric optimization. However, metamodel based optimization is not well suited for material systems with highly nonlinear behavior, disjoint design space, limited data and high experiment/simulation cost; a common scenario in material science. In contrast, Bayesian Optimization (BO) [12, 13] has emerged as a viable proposition in material design. BO adaptively samples new designs conditioned on information available from existing data, balancing exploration and exploitation to efficiently locate the global optimum. Balachandran et al. [14] used this adaptive sampling strategy to design $M_2AX$ compounds with desired elastic properties using orbital radii of individual components as design variables. Their work highlighted the importance of incorporating prediction uncertainties in the sampling procedure and noted that purely exploitative strategies often result in suboptimal outcomes. Li et al. [15] devised an adaptive experimental optimization (AEO) framework to develop a novel fluid processing platform for synthesis of short polymer fibers. The AEO framework incorporates material and process related parameters to optimize a set of qualitative and quantitative objectives. All design variables are quantitative and related to process optimization for a single combination of polymer and solvent. Similar examples of accelerated material design using BO have been reported by others [16, 17]. However, all existing applications of BO considered thus far have only quantitative or qualitative variables; while mixed variable problems containing both qualitative and quantitative variables is common in material design. Choice of constituents in any material system can be treated as qualitative variables, while microstructure descriptors, processing and operating parameters (temperature, RPM, wavelength etc.) manifest as quantitative variables. For example, nanocomposite design involves selecting the optimal combination of qualitative (polymer, nanoparticle, surface modification) and quantitative (microstructure descriptors) variables.

In this article, we present a data-centric Bayesian Optimization framework for material design and innovation. The guiding hypothesis is that a Bayesian inference approach can effectively model knowledge contained in an available dataset as a prior and guide "on-demand" computer simulations and physical experiments to accelerate the search of optimal material designs. Therefore, the framework is flexible to incorporate data generated by experiments as well as simulations or machine learning. We demonstrate how Latent Variable Gaussian Processes, a novel GP modelling strategy for mixed variable problems with inbuilt uncertainty estimation, plays a critical role in Bayesian Optimization to efficiently navigate complex, non-linear design space. Then, LVGP based BO is extended to multi-objective optimization, a common scenario in design of multi-functional material systems. The efficacy of the proposed framework is demonstrated through a design case study focused on identifying the optimal composition and microstructure for insulating nanocomposites.

## 2. DATA-CENTRIC MIXED-VARIABLE BAYESIAN OPTIMIZATION (BO) FRAMEWORK FOR MATERIALS DESIGN

In this section, we present the data-centric material design framework and discuss the two driving concepts of BO – surrogate model and acquisition functions.

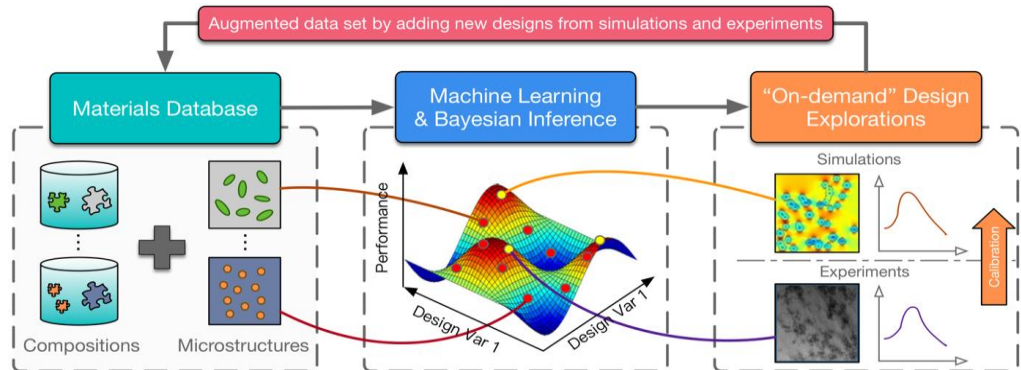

Figure 1: Bayesian optimization approach treats the existing data set as prior knowledge, chooses new samples, and builds machine learning models using curated and new experiment/simulation data to capture p-s-p relations for optimization.

## 2.1 Data-Centric Material Design Framework and Concept of BO

Fig. 1 shows the proposed framework, integrating curated material databases with material property simulations and machine learning. The framework is initiated from a **materials database** of curated experimental and simulated data describing material properties with appropriate attributes. For example, the NanoMine [5] system we developed contains data describing mechanical, electrical and optical properties of nanocomposites – labelled by composition (polymer, nanoparticle, surface chemistry, nanoparticle weight ratio), microstructure, processing conditions and associated metadata (source of data, processing equipment used, etc.). Based on PSP relationships, some of these material characteristics are known to influence material properties and are treated as design variables in BO. These characteristics maybe quantitative (e.g., microstructure descriptors) or qualitative (e.g., type of nanoparticle or polymer, type of surface treatment). Material database can be used to train **machine learning** models which predict material properties from the corresponding design variables. Since the database contains only a small fraction of points in a vast design space, the machine learning model possesses interpolation uncertainty, which varies from point to point and plays an important role in BO-based **"on-demand" design explorations**. Using predictions and uncertainty quantification of a machine learning model, **Bayesian inference** determines the design that shows the most "potential" for improvement in a given material property. There are several metrics, commonly known as **acquisition functions**, for evaluating "potential" improvement. After a promising design is identified by the acquisition function, its corresponding material property is evaluated using either simulation or experiment. Once property evaluation is complete, the sample design is added to the database and the above steps will be repeated until the termination criteria is satisfied based on available computational and experimental resources. It should be noted that as a part of the Bayesian inference framework, experimental data can also be used for calibrating unknown model parameters in simulation models. For example, finite element simulations for prediction of dielectric properties in nanocomposites requires calibration of change in polymer mobility at the polymer-nanoparticle interphase (the region surrounding nanoparticles) as characterized by the shift in dielectric loss peaks [18] obtained from the experimental dielectric spectra.

Let us represent the general global optimization problem as

$$x^* = \underset{x \in X}{\mathrm{argmin}}\, f(x), \tag{1}$$

where $x$ is the input design variable vector in the input space $X$ and $f(\cdot)$ is the objective function. Based on the data collected about $f(\cdot)$, BO uses statistical models to model the true $f(\cdot)$ function and produce fast statistical predictions of $f(x)$ at any given input $x$. The statistical model must be flexible to tackle situations where $x$ is qualitative, quantitative or a combination of both. We adopt Gaussian process (GP) models as the statistical model in our BO framework as GP models can flexibly model highly nonlinear behavior with a small amount of hyperparameters and conveniently provides analytical solutions for its prediction uncertainty.

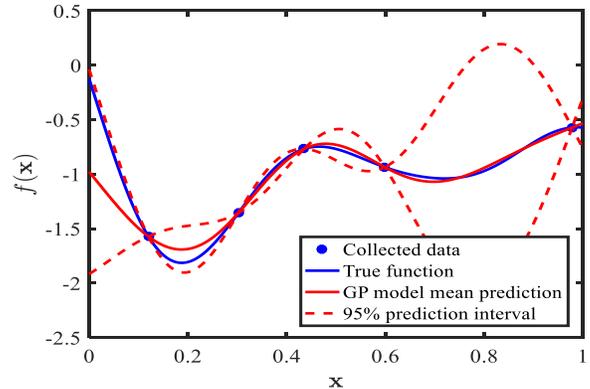

Figure 2: 1D example of a GP model fitted to collected data of $f(\cdot)$

Fig. 2 is a 1D example of a GP model fitted to the collected data of $f(\cdot)$. At each input $x$, the output $f(x)$ is regarded as a normally distributed random variable and the GP model predicts its mean and variance. The 95% prediction interval in the figure reflects the confidence bounds of the prediction [19, 20].

The **acquisition functions** (aka utility functions, infill functions) in BO are functions that quantify the benefit of evaluating $f(x)$ at any input $x$ in the next optimization iteration based on the statistical predictions of $f(x)$. Fig. 3 illustrates profiles of a few widely used acquisition functions for the 1D example problem in Fig 2. Plotted acquisition functions are expected improvement (EI) [21], probability of improvement (PI) [22], upper/lower confidence bound (UCB) [23] and knowledge gradient (KG) [24]. In general, an acquisition function aims for either exploitation, exploration, or both of a design space. Here, **exploitation** means evaluating at input $x$ where prediction of $f(x)$ has high cumulative probability of being better than the current best solution from the perspective of optimizing a design objective. On the other hand, **exploration** implies evaluating at input $x$ where prediction of $f(x)$ has large uncertainty, which warrants the chance that either $x$ is a potential global optimum or the evaluation of $f(x)$ leads to the discovery of the global optimum near $x$. On many occasions, exploitation and exploration are quite contradictory goals to achieve. Therefore, acquisition functions typically strike a balance between the two.

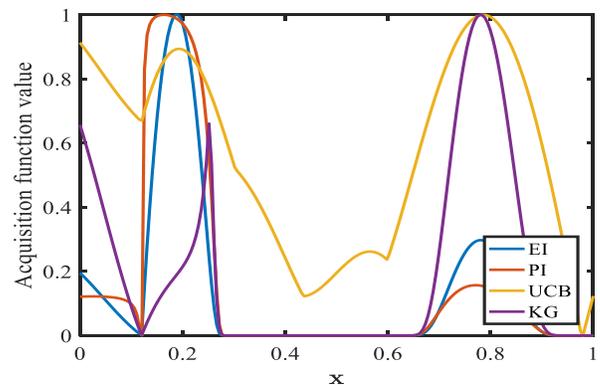

Figure 3: Different acquisition function profiles in a 1D example problem, based on the statistical predictions in Fig.2. The value of the acquisition functions is a measure of the benefit of evaluating f(·) at that input.

## 2.2 Latent Variable GP Modelling for Mixed-Variable Problems

The standard GP methods were developed under the premise that all input variables are quantitative, which does not hold in many real engineering applications. We recently proposed a latent variable Gaussian processes (LVGP) [25] modeling method that maps the levels of the qualitative factor(s) to a set of numerical values for some latent quantitative variable(s). The latent variables transform the underlying high dimensional physical attributes associated with the categorical variables into the latent-variable space and quantify the "distances" between samples. As a by-product, the latent variable mapping of the qualitative factors provides an inherent ordering and structure for the levels of the factor(s), which leads to substantial insight into the effects of the qualitative factors. To avoid misunderstanding, note that the latent variables are only used internally inside LVGP models. When LVGP models are used for predictions, they still take mixed-variable inputs in the original mixed-variable input spaces.

To describe the LVGP approach, the input variables are denoted as $w = (x, t)$, where $x = (x_1, x_2, \ldots, x_p)$ represents $p$ quantitative variables and $t = (t_1, t_2, \ldots, t_q)$ is the vector of $q$ qualitative variables. With $i = 1, 2, \ldots, q$, the qualitative variable $t_i$ has $m_i$ levels $\{l_1^{(i)}, l_2^{(i)}, \ldots, l_{m_i}^{(i)}\}$. The output variable is denoted as $y$, and a set of data points of input-output pairs are noted as $\{(w_1, y_1), \ldots, (w_N, y_N)\}$. Consider the GP model
$$Y(\cdot) = \mu + G(\cdot), \quad (2)$$
where $\mu$ is the constant prior mean, and $G(\cdot)$ is a zero-mean GP with covariance function $k(\cdot, \cdot) = \sigma^2 r(\cdot, \cdot | \varphi)$. $\sigma^2$ is the prior variance of the GP, and $r(\cdot, \cdot | \varphi)$ is the correlation function parameterized with $\varphi$. The true model $y(\cdot)$ is regarded as a realization of the GP $Y(\cdot)$. Once the form of the correlation function $r(\cdot, \cdot | \varphi)$ is specified, the hyperparameters $(\mu, \sigma^2, \varphi)$ can be estimated through maximum likelihood estimation (MLE) or other principles such as minimizing cross-validation errors. If the independent variables of the correction function $r(\cdot, \cdot | \varphi)$ are only the continuous variables $x$, one can use the common Gaussian correlation function
$$r(x, x'|\varphi) = exp\left\{-\sum_{i=1}^{p} \varphi_i(x_i - x_i')^2\right\}, \quad (3)$$
which quantifies the correlation between $G(x)$ and $G(x')$ for any input locations $x = (x_1, \ldots, x_p)$ and $x' = (x_1', \ldots, x_p')$ based on their 2-norm distance scaled by $\varphi$. However, in the mixed-variable problem, it is not straightforward to incorporate the qualitative variable $t$ in such a correlation function, as it makes no sense to compute $t_i - t_i'$ as the difference between levels. The LVGP method handles this by mapping the qualitative variables $t$ to quantitative ones.

In the LVGP method, the $m_i$ levels of the qualitative variable $t_i$ are mapped to $m_i$ latent numerical vectors $\{z^{(i)}(l_1^{(i)}), \ldots, z^{(i)}(l_{m_i}^{(i)})\}$ of a latent variable $z^{(i)} \in \mathbb{R}^d$, where $d$ is the dimensionality of $z^{(i)}$. A modeler is free to choose the value of $d$ as a modeling parameter, although setting $d = 2$ has been shown to be advisable for most problems. The original mixed-type input variables $w = (x, t)$ are thus mapped to purely continuous variables $(x, z^{(1)}(t_1), \ldots, z^{(q)}(t_q))$. A correlation function like Eq. 3 can be subsequently constructed as
$$r(w, w'|\varphi, Z) = exp\left\{-\sum_{i=1}^{p} \varphi_i(x_i - x_i')^2 -\sum_{i=1}^{q} \|z^{(i)}(t_i) - z^{(i)}(t_i')\|_2^2\right\}, \quad (4)$$
where $Z$ is the collection of all the latent parameters denoted by $\{z^{(1)}(l_1^{(1)}), \ldots, z^{(1)}(l_{m_1}^{(1)}), z^{(2)}(l_1^{(2)}), \ldots, z^{(q)}(l_{m_q}^{(q)})\}$.

With the correlation structure in Eq. 4, and following Eq. 2, the log-likelihood for the given dataset $\{(w_1, y_1), \ldots, (w_N, y_N)\}$ is
$$L(\mu, \sigma^2, \varphi, Z) = -\frac{N}{2}\ln(2\pi\sigma^2) - \frac{1}{2}\ln|R(\varphi, Z)| - \frac{1}{2\sigma^2}(y - \mu\mathbf{1})^T R(\varphi, Z)^{-1}(y - \mu\mathbf{1}), \quad (5)$$
where $y$ is the column vector $(y_1, \ldots, y_N)^T$, $\mathbf{1}$ is the $N$-by-1 vector of ones, and $R(\varphi, Z)$ is an $N$-by-$N$ matrix with $(i,j)^{th}$ element being $R_{ij} = r(w_i, w_j | \varphi, Z)$ for $i, j = 1, \ldots, N$. The MLE values of the hyperparameters $(\mu, \sigma^2, \varphi, Z)$ are obtained as
$$[\hat{\mu}, \widehat{\sigma^2}, \hat{\varphi}, \hat{Z}] = arg \max_{\mu, \sigma^2, \varphi, Z} L(\mu, \sigma^2, \varphi, Z). \quad (6)$$

For estimating the hyperparameters, the latent vector $z^{(i)}(l_1^{(i)})$ corresponding to the first level $l_1^{(i)}$ is fixed at zero to avoid indeterminacy issue. This is because the correlation quantified by Eq. 4 only depends on the relative distances between $z(t)$ and $z(t')$ but not their respective absolute values. When $d \geq 2$, some other elements in the collection of latent vectors $\{z^{(i)}(l_1^{(i)}), ..., z^{(i)}(l_{m_i}^{(i)})\}$ need to be set to zero to avoid indeterminacy. In the $d = 2$ case, the second element of $z^{(i)}(l_2^{(i)})$ needs to be set to zero, as is illustrated in Fig 4. The total number of latent parameters that are subject to optimization is thus $\sum_{i=1}^{q}(2m_i - 3)$.

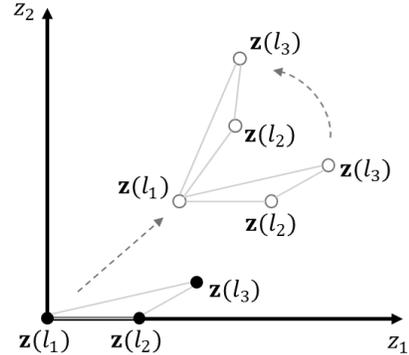

Figure 4: Indeterminacy caused by translation and rotation when $d = 2$. Fixing the continuous variable values, the three different configurations for the mapped latent values $\{z(l_1), z(l_2), z(l_3)\}$ have the same pairwise distances and hence the same correlation values. Index $i$ is omitted for simiplicity.

After obtaining the estimates of the hyperparameters, the mean prediction for $y(w_{(n)})$ and the associated mean squared error (MSE) (for uncertainty prediction) are

$$\hat{y}(w_{(n)}) = \hat{\mu} + r^T(w_{(n)})R^{-1}(y - \hat{\mu}\mathbf{1}), \quad (7)$$

$$MSE[\hat{y}(w_{(n)})] = \hat{\sigma}^2[r(w_{(n)}, w_{(n)}) \\ - r^T(w_{(n)})R^{-1}r(w_{(n)}) \\ + W^2(\mathbf{1}^T R^{-1}\mathbf{1})^{-1}], \quad (8)$$

where $W = 1 - \mathbf{1}^T R^{-1} r(w_{(n)})$ and $r(w_{(n)})$ is a $N$-by-1 vector whose $i^{th}$ element is $r(w_i, w_{(n)})$.

### 2.3 Multi-Objective Bayesian Optimization

A multi-objective BO approach is needed because material design often involves targets for multiple properties. The general multi-objective optimization problem can be formulated as

$$\min_{x \in X}\{y_1(x), y_2(x), ..., y_s(x)\}, \quad (9)$$

where $x$ is the design input, $X$ is the design space, $s$ is the number of objective functions, and $\{y_1(\cdot), y_2(\cdot), ..., y_s(\cdot)\}$ is the set of the objective functions that share the same design inputs. The solution to this problem is a so-called Pareto set (aka Pareto front or Pareto frontier) consisting of design points that achieve Pareto optimality [26]. To identify the Pareto frontier for Eq. 9 numerically, the objective functions are evaluated at a certain number of design inputs. Of all the evaluated design points, one selects the set of design points that are not dominated by any other ones [26]. Here, a design point $x$ is not dominated by another one $x'$ if there exists at least one $i \in \{1,2, ..., s\}$ such that $y_i(x) < y_i(x')$. This set of design points is regarded as a representation of the true Pareto set.

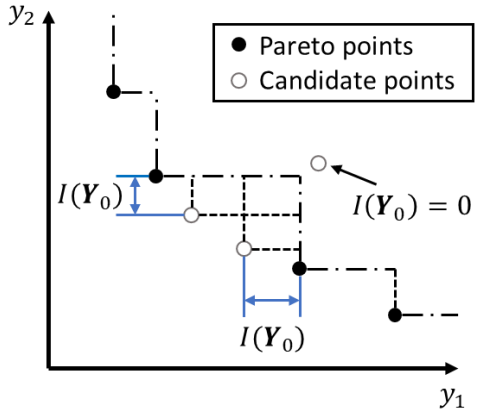

Figure 5: Example of the values of $I(Y_0(x_0))$. The depiction is in 2D output space.

Several BO methods have been proposed in the literature for such a multi-objective setting. In our implementation, we choose the widely-used expected maximin improvement (EMI) formulation [27], which we describe as follows. Let the current Pareto set be composed of input set $P_X = \{x_1, x_2, ..., x_k\}$ and output set $P_Y = \{y_1, y_2, ..., y_k\}$, where $k$ is the number of points in the Pareto set and $y_i = [y_1(x_i), y_2(x_i), ..., y_s(x_i)]^T, i = 1, 2, ..., k$. For any given new input $x_0$, the corresponding output is predicted by the uncertainty quantification model as $Y_0(x_0) = [Y_1(x_0), Y_2(x_0), ..., Y_s(x_0)]^T$, where $Y_j(x_0), j = 1, 2, ..., s$ is a random variable. To quantify how much the random outputs $Y_0(x_0)$ would improve the current Pareto set, we use the minimax improvement metric

$$I(Y_0(x_0)) = \min_{x_i \in P_X}\left\{max\left(\{y_j(x_i) - Y_j(x_0)\}_{j=1}^{s} \\ \cup \{0\}\right)\right\}. \quad (10)$$

The larger the value of $I(Y_0(x_0))$ is, the more improvement the output $Y_0(x_0)$ is considered to make.

With this formula, if the output $Y_0(x_0)$ would be dominated by at least one point in the current Pareto set, then $I(Y_0(x_0)) = 0$, which means no improvement. Otherwise, $I(Y_0(x_0))$ would be a positive value quantifying the improvement. The value of $I(Y_0(x_0))$ is illustrated by a 2D example case in Fig. 5, with one of the candidate points being $I(Y_0) = 0$ and the other two points with a positive value $I(Y_0)$. The criterion for choosing the new evaluation input $x_0^*$ is to maximize the EMI given in Eq. 10, i.e.,

$$x_0^* = \underset{x_0 \in X}{argmax} E(I(Y_0(x_0))). \quad (11)$$

When the original problem (Eq. 9) has mixed-variable input space $X$, Eq. 11 is a mixed-variable optimization problem. To solve Eq. 11, we use a zero-order optimization strategy, where we generate a large set of candidate points in the input space, and then choose the one with the largest EMI as $x_0^*$. For evaluating the expectation in Eq. 11, we use Monte Carlo simulation, as the analytical formula for EMI is too complex for $s \geq 3$, which is the case for our application problems.

### 3. CASE STUDY: CONCURRENT COMPOSITION AND MICROSTRUCTURE DESIGN FOR INSULATING NANOCOMPOSITES

Polymer nanocomposites are ideal candidates for insulating materials with potential application in high voltage rotating machines [28]. Three major electrical properties that determine suitability for this application are breakdown strength, dielectric permittivity and dielectric loss. Breakdown strength ($U_d$) is the minimum voltage at which current flows through an insulating material. Dielectric permittivity ($\epsilon$) characterizes the degree of electrical polarization experienced by material while dielectric loss ($tan\delta$) is a measure of the amount of heat generated under an alternating electric field. Ideally, one would want high $U_d$, low $\epsilon$ and low $tan\delta$ but tradeoffs between $U_d$ Vs $\epsilon$ and $\epsilon$ Vs $tan\delta$ have been observed [29, 30]. Automation of polymer nanocomposite design is considerably less developed than that for alloys due to challenges associated with searching the vast design space defined by the almost infinite combinations of polymers, nanoparticle, surface chemistries, microstructure morphology, and processing conditions.

Fig. 6 depicts the mixed-variable BO framework customized for case study, indicating the various modules involved and information flow between them. Our investigations are initiated from a materials database (**Module 1**) comprising nanocomposite samples with varying compositions, corresponding microstructures and measurement of dielectric properties. We consider nanocomposites with silica nanoparticles (aka filler) dispersed in two types of polymers - polystyrene (PS) and polymethylmethacrylate (PMMA). It is a common practice to enhance interphase properties by appending functional groups to the surface of nanoparticles, a process known as surface modification. We consider silica nanoparticles with three surface modifications in this study – Chloro-, Amino- and Octyl-silanes. Samples corresponding to all six polymer and surface modification combinations were prepared. Nanoparticle dispersion, a key microstructure descriptor influenced by choice of polymer, surface modification and processing conditions is quantified from TEM images using the Spectral Density Function (SDF) along with nanoparticle volume fraction. The identified range of these microstructure descriptors will be used as bounds in the design process.

Our database also contains experimental measurements of all three dielectric properties, to be used for calibrating the nanoparticle-polymer interphase parameters in the finite element analysis (FEA) model as well as training machine learning models for the breakdown strength. These properties are known to be influenced by material composition (choice of filler, polymer, surface modification), filler volume fraction and dispersion. Dielectric permittivity $\epsilon$ and loss $tan\delta$ are evaluated using FEA (**Module 3**), where interphase properties are characterized by a shift in the nanocomposite properties w.r.t pure polymer properties and obtained by calibration (**Module 2**). In **Module 3** SDF based microstructure reconstruction is used to generate 2D Representative Volume Elements (RVEs) with desired filler volume fraction and dispersion; which is evaluated using FEA. Since microstructure reconstruction using SDF is stochastic and involves uncertainties, we generate three reconstructions and take their average $\epsilon$ and $tan\delta$ to calculate the objective function for each design. **Module 3** represents the most computationally intensive module in our framework, requiring several minutes of simulations on a 16 core Intel Xeon® 2.4 GHz CPU with 192 GB RAM. On the other hand, **Module 4** is an empirical machine learning model employing Random Forrest technique, trained on experimental data present in nanocomposite database, to predict the breakdown strength $U_d$ as a function of material design variables.

With bounds for design variables identified and models to predict dielectric properties, our study progresses to BO for designing insulating nanocomposites (**Module 5**). BO is initiated with a small set of training data and adaptively samples subsequent designs to approach the global optimum. We use LVGP with two-dimensional latent variables for each qualitative variable, as the surrogate model. Its in-built uncertainty quantification is leveraged for performing Bayesian Inference using the Expected Improvement acquisition criterion. At each iteration, the LVGP model is updated with a new design whose dielectric properties are evaluated using Modules 3 & 4. We present single and multi-objective optimization formulations for the nanocomposite design problem and discuss performance of LVGP-BO framework in both scenarios. In the following subsections, we describe each module in detail, followed by results.

### 3.1 Nanocomposite Preparation and Dielectric Property Measurement (Module 1)

Silica nanoparticles (diameter 14 nm) in methyl ethyl ketone were procured from Nissan Inc. The surface of the nanoparticles was modified using three monofunctional silane coupling agents:

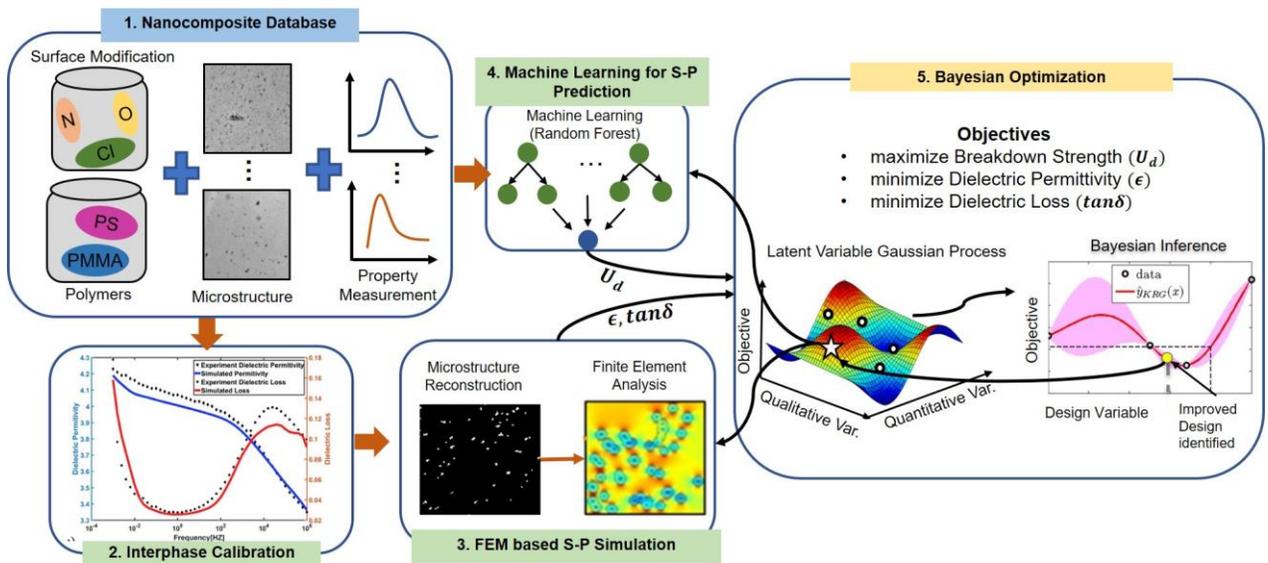

Figure 6: Bayesian Optimization framework for insulating polymer nanocomposites design

aminopropyledimethylethoxysilane (Amino), chloropropyledimethylethoxysilane (Chloro) and octyldimethylmethoxysilane (Octyl), from Gelest Inc. Polystyrene (PS) from Goodfellow Corporation and polymethylmethacrylate (PMMA) from Scientific Polymer Products Incorporated were used as the polymer. Nanocomposites were prepared in a Thermo Haake Minilab, co-rotating twin screw extruder. The mixing was carried out at 200°C for the PMMA nanocomposites and 150°C for the PS nanocomposites. Mixing parameters such as screw speed and specific energy input were varied to obtain a range of different dispersion states. A JEOL 2010 transmission electron microscope (TEM) was used to characterize the dispersion state of the nanocomposites. For each sample, 30 images were collected from different sections to obtain a representation of the overall dispersion state of the nanocomposites. The TEM images were binarized using a Niblack algorithm [31]. Dielectric spectroscopy measurements to ascertain $\epsilon$ and $tan\delta$ were carried out in a Novocontrol Alpha Analyzer. All nanocomposite samples were also subjected to dielectric breakdown testing. Hot pressed samples were placed in a ball-plane electrode setup [32]. The testing apparatus was kept in an oil bath to reduce the incidences of flash over and the voltage ramp-rate was fixed at 500 V/s. For each nanocomposite sample, 30 specimens were tested and the voltage at breakdown for each sample was recorded. The observed values were then fit to a Weibull function to obtain the dielectric breakdown strength, which is defined as the electric field at which the probability of breakdown was 63.2%.

### 3.2 Microstructure Characterization and Reconstruction (Module 1 & 3)

Spectral Density Function (SDF) is a frequency domain microstructure representation capable of capturing spatial correlations of complex heterogeneous materials. Mathematically, SDF $\rho(k)$ can be evaluated as:

$$\rho(\boldsymbol{k}) = |\mathcal{F}\{\mathcal{M}\}|^2, \qquad (12)$$

where $\mathcal{M}$ is the binarized microstructure, $\mathcal{F}(.)$ is the Fourier Transform operator and $\boldsymbol{k}$ is the frequency vector. For isotropic microstructures, SDF can be radially averaged about zero frequency such that the frequency vector $\boldsymbol{k}$ is reduced to a scalar $k$; making SDF a one-dimensional function of frequency. Although it is known to be the Fourier Transform of Two-point Autocorrelation function and hence encapsulates equivalent morphological information, Yu et al. [9] have shown that SDF is a more convenient representation to parametrize and design microstructures. These features are also evident from the analysis of nanocomposite microstructures. After binarizing TEM images using Niblack Algorithm [31] and assuming isotropy, SDF was evaluated using Eq. 12. We noticed that SDF of all microstructures approximately follows an exponential distribution that can be parametrized with two variables – shape parameter $\alpha$ and scale parameter $\theta$:

$$\rho(k) = \alpha * \exp\left(-\frac{k}{\theta}\right). \qquad (13)$$

A total of 1719 TEM images (approx. 30 images per sample) were characterized using SDF and parameters $\alpha$ and $\theta$ were ascertained by curve fitting using Eq. 13. The average $R^2$ value for fitting was 0.91. Fig. 7 shows three microstructures along with their one-dimensional SDF and curve fitting. Filler dispersion increases from Fig. 7a-c and is reflected in a slower decay rate of SDF which can be quantified by $\theta$. It was noticed that $\alpha$ varies in a very small interval [0.39, 1.84] and has very little influence on SDF profile. On the other hand, scale parameter $\theta$ varies between [0.62, 6.55], changing the rate of decay of SDF and consequently characterizing the dispersion of filler aggregates. Thus, we will consider $\theta$ as a microstructure design variable and fix $\alpha$ to its mean value 1.1. Additionally, filler composition is represented by its volume fraction (VF), computed as the fraction of white pixels in binarized image, and is found to be in [0.7% , 8.3%]. The range of $\theta$ and VF identified here will be used to define bounds for these variables in design formulation.

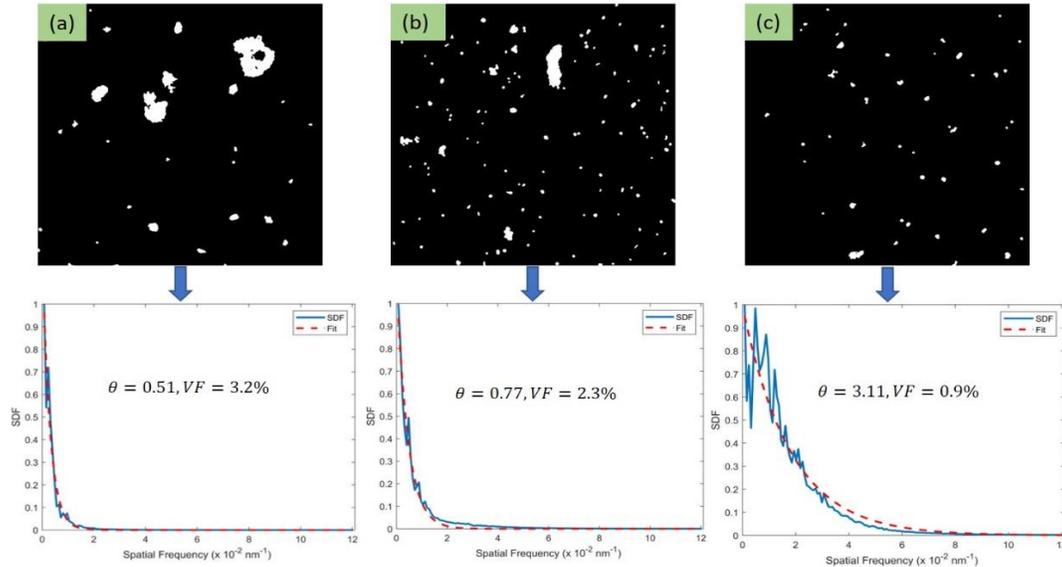

Figure 7: Three representative microstructure images with varying dispersions and their SDF (blue curve) and corresponding curve fit using Eq. 13(red dashed curve). Design Variables $VF$ and $\theta$ for each image shown in inset

Table 1: Summary of design variables used in case study

| Variable | Type | Range/Levels |
|---|---|---|
| Polymer Type ($P$) | Qualitative | $\{PMMA, PS\}$ |
| Surface Modification Type ($S$) | Qualitative | $\{Chloro, Octyl, Amino\}$ |
| Filler Volume Fraction ($VF$) | Quantitative | $[0.7\%, 8.3\%]$ |
| Filler Dispersion ($\theta$) | Quantitative | $[0.62, 6.55]$ |

Microstructure reconstruction is an integral part of the material design framework, since at each iteration of optimization, the material properties must be evaluated for the new microstructure. Here we adopt the analytical Cahn's Scheme [33] for reconstruction.

### 3.3 Interphase Calibration and Finite Element Analysis for Dielectric Permittivity and Loss (Module 2 & 3)

Finite element (FE) models have been developed to simulate the continuum level properties for polymer nanocomposites [34-36]. One of the challenges in structure-property relationship of nanocomposites is modeling the interphase. Since direct measurement of interphase properties, especially in situ, is limited experimentally, one approach is to calculate the interphase properties inversely. Using a FEA model with known nanoparticle and polymer properties, the interphase properties can be tuned until the predicted properties match the experimental measurement of the bulk composite properties [37, 38]. We developed an automated optimization-based method to solve this inverse calibration problem, details of which can be found in Wang et al. [18]. This method helps identify the dielectric shifting factors that will optimally match the experimental data through adaptive optimization. For the case study, interphase properties for all six polymer-surface modification combinations are calibrated using this adaptive optimization strategy (Module 2) and later used in FEA model (Module 3).

### 3.4 Machine Learning for Breakdown Strength Prediction (Module 4)

Dielectric breakdown of nanocomposites is a complex phenomenon and requires atomistic scale simulations to decode the complex interactions occurring in the interphase. Our investigation in this direction are ongoing. Therefore, in this case study, we use a random forest [39] model trained on experimental data for rapid evaluation of $U_d$ as a function of material design variables during optimization. Random forest technique was chosen due to its ability of handling mixed variables, superior computational efficiency and minimal possibility of overfitting. Training data comprised $U_d$ measurement (expressed in kV/mm) of 52 samples at 60 Hz. The features used for predicting $U_d$ are the two qualitative (polymer type, surface modification type) two quantitative ($VF$ and $\theta$) design variables. The trained random forest with 1000 trees has $R^2 = 0.9$.

### 3.5 Material Design Formulation for Mixed variable BO (Module 5)

Our goal is to identify nanocomposites with high $U_d$, low $\epsilon$ and low $tan\delta$ suitable for electrical insulation. The design space consists of four variables, two qualitative and two quantitative, as summarized in Table 1. Choice of polymer and surface modification are qualitative variables with two (PS, PMMA) and three (Octyl, Chloro, Amino) levels respectively. Dispersion and volume fraction are quantitative variables with their bounds identified using SDF in Section 3.2. We present both single and multi-objective BO strategies for this case study, using the same set of design variables with different objective formulations. For single objective BO, we formulate an objective function that weighs all three normalized properties (indicated by *) equally and adds (subtracts) each property depending on whether it needs to be minimized (maximized):

$$\min_{s \in S, p \in P, m \in M} tan\delta^* + \epsilon^* - U_d^*$$
$$S: \{Chloro, Octyl, Amino\}$$
$$P: \{PMMA, PS\} \quad (14)$$
$$M: microstructures \ with \ 0.7\% \leq VF$$
$$\leq 8.3\%, 0.62 \leq \theta \leq 6.55,$$

where objective is to be minimized over a design space consisting of all possible combinations of surface modification ($S$), polymers ($P$) and microstructures ($M$). In contrast, multi-objective aims to find candidate designs lying on the Pareto Frontier – a characteristic boundary comprising designs where no objective can be improved without deterioration of others. With three dielectric properties of interest, we define a multi-objective optimization problem as follows:

$$\min_{s \in S, p \in P, m \in M} tan\delta, \epsilon, -U_d,$$
$$S: \{Chloro, Octyl, Amino\}$$
$$P: \{PMMA, PS\} \quad (15)$$
$$M: microstructures \ with \ 0.7\% \leq VF$$
$$\leq 8.3\%, 0.62 \leq \theta \leq 6.55,$$

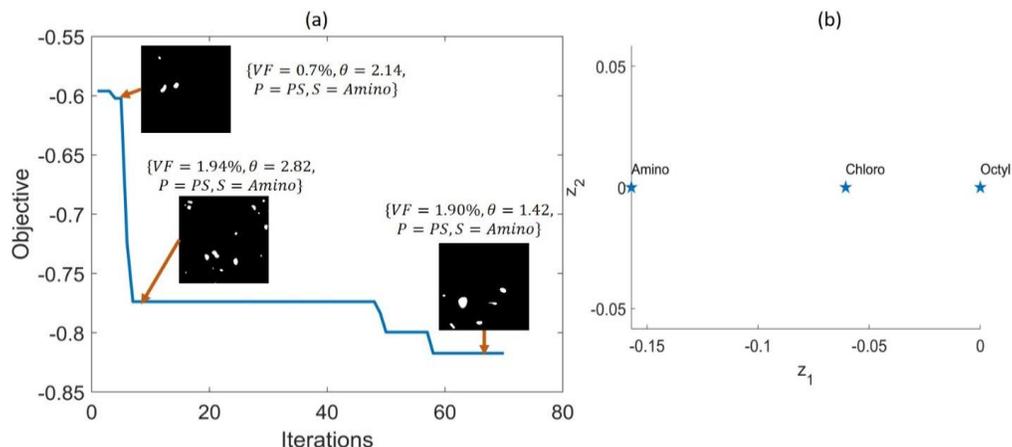

Figure 8: (a)Optimization history for single objective BO that converged to objective = -0.817 along with three designs evaluated in the process (b) Latent variable space for surface modification ($S$) variable

In this case, we do not normalize dielectric properties and all variables have their usual meaning.

### 3.6 Results and Discussion

We performed 70 & 120 iterations of BO for single and multi-objective formulations respectively, as specified by Eq. 14-15. Each BO is initiated with 30 random initial samples where quantitative variables $\{VF, \theta\}$ are generated by Latin Hypercube Design and qualitative variables polymer & surface modification type are sampled uniformly.

### 3.6.1 Results from Single Objective Bayesian Optimization

We performed ten replicates of single objective BO; each replicate initiated with 30 random samples. We observed that all replicates converge to optimal design with objective value of $-0.798 \pm 0.009$. The best solution identified was -0.817, which corresponds to design $\{VF = 1.9\%, \theta = 1.42, P = PS, S = Amino\}$ with material properties $tan\delta = 0.0013$, $\epsilon = 2.108$ and $U_d = 132.421 \frac{kV}{mm}$. Fig. 8a shows optimization history for one replicate and depicts evolution of design during optimization. We observe that amino-modified Silica nanoparticles in PS with low filler volume fraction and dispersion is ideal to meet our requirements of high $U_d$, low $tan\delta$ and $\epsilon$. These findings are consistent with our previous investigations that found $tan\delta$ and $\epsilon$ increase with dispersion. An added benefit of using LVGP is the ability to visualize latent space and draw insights about behavior of various levels for each qualitative variable. Fig. 8b shows the latent variable space for $S$- the three-level qualitative variable representing surface modification. We notice that amino surface modification is positioned relatively further away from other two – suggesting that its effect on dielectric properties is different. Also, we notice in Fig. 8b that all three levels are positioned along the $z_1$ axis with little variation in $z_2$; suggesting that the three surface modifications depend on a single latent factor and the use of two-dimensional latent variable representation is sufficient. We believe this latent factor may be the interfacial energy descriptor described by Hassinger et al. [40]. The ordering of these three levels, Octyl-Chloro-Amino (or vice-versa), also matches the order of their corresponding interfacial energy descriptor.

To demonstrate the efficacy of BO in identifying optimal designs for problems with limited computational budget, we compare its performance against Genetic Algorithm (GA) [41]. MATLAB's implementation of GA for mixed integer optimization was used in this study and applied to problem formulation defined by Eq. 14. For a fair comparison with BO, GA was configured to terminate after 100 objective function evaluations (10 generations with population size of nine). Fig. 9 compares the optimal designs identified by 10 replicates of GA versus BO. Each BO replicate was initiated with 30 random samples. We see that regardless of initial samples provided, BO can consistently converge to the optimum design with low variability whilst GA is highly susceptible to the initial population that's usually generated randomly. This shows that the BO strategy of utilizing GP model uncertainty quantification to intelligently select new designs for evaluation and improve surrogate LVGP model makes it robust & faster at approaching

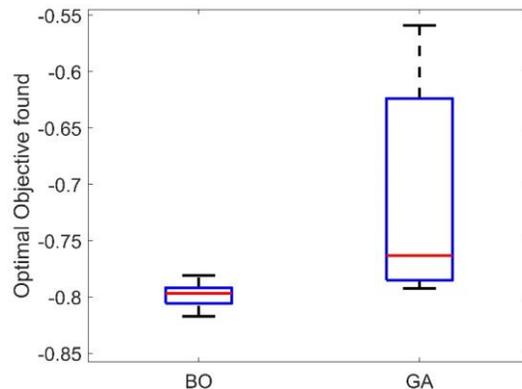

Figure 9: Comparison of 10 replicates of BO and GA for single objective optimization

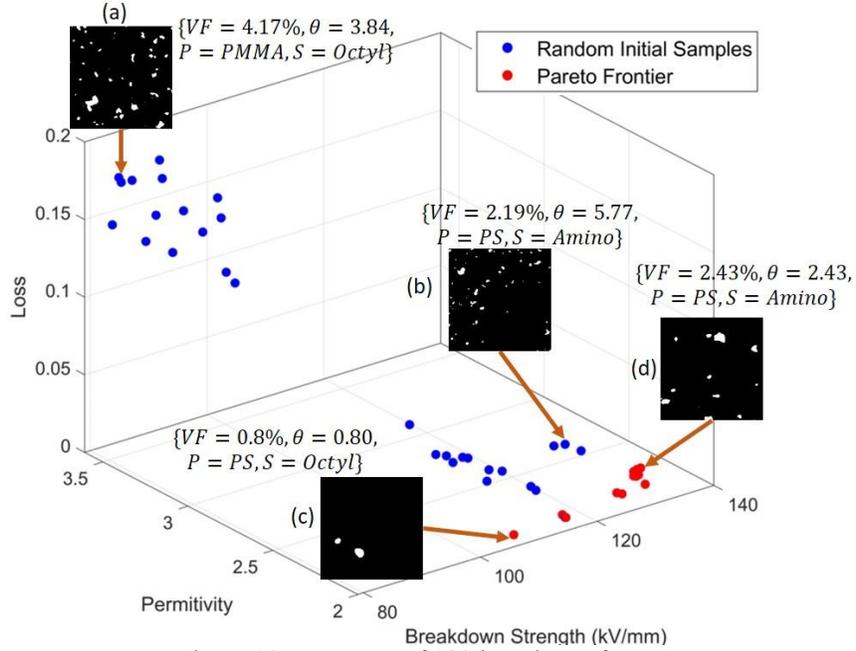
Figure 10: Summary of 120 iterations of MOBO

global optimum compared to other algorithms that ignore this information.

### 3.6.2 Results from Multi-Objective Bayesian Optimization (MOBO)

120 iterations of MOBO were performed starting with 30 random initial samples. Fig. 10 plots the random samples and 11 designs that were identified on the Pareto front. A noticeable feature in this plot is that the initial samples create two clusters corresponding to two polymers under consideration. The cluster located in the low $U_d$, high $tan\delta$ and $\epsilon$ region (top left corner in Fig. 10) exclusively contains PMMA based samples. This is consistent with our knowledge of dielectric behavior of PMMA. On the other hand, PS based samples have higher $U_d$, lower $tan\delta$ and $\epsilon$; suggesting that they are better suited for electrical insulation application as compared to PMMA samples. This is also reflected in the fact that all 11 designs identified on the Pareto Front are exclusively PS based. Notice that Pareto Front obtained by MOBO shows significant improvement w.r.t random initial samples and thus underlines the capability of uncertainty driven MOBO to locate improved designs.

The designs on the Pareto Frontier show large variations for $U_d$ as compared to those of $tan\delta$ and $\epsilon$. For example, designs (c) and (d) in Fig. 10 has properties $\{U_d = 107.83 \frac{kV}{mm}, \epsilon = 2.07 \ \& \ tan\delta = 0.0012\}$ and $\{U_d = 135.05 \frac{kV}{mm}, \epsilon = 2.26 \ \& \ tan\delta = 0.0020\}$ respectively. Table 2 lists all Pareto designs and their dielectric properties. It suggests that designs with octyl surface modification have lower $tan\delta$, $\epsilon$ and $U_d$ while amino surface modification induces higher $tan\delta$, $\epsilon$ and $U_d$. The behavior of designs with chloro surface modification is intermediate; as suggested by its relative positioning in latent space in Fig. 8b. Similarly, high (low) filler $VF$ and $\theta$ leads to high (low) $\tan\delta$, $\epsilon$ and $U_d$. Thus, we see a tradeoff between the three properties of interest. Which point in the Pareto front should be chosen as the design solution will depend on the designer's preference function, to be created based on the application, how the material is deployed, and device level performance.

Table 2: Pareto points obtained by MOBO

| P | S | VF(%) | θ | $U_d$ (kV/mm) | $\epsilon$ | $tan\delta$ |
|---|---|---|---|---|---|---|
| PS | Octyl | 0.884 | 0.803 | 107.835 | 2.072 | 0.0012 |
| PS | Octyl | 1.13 | 1.417 | 117.165 | 2.107 | 0.0014 |
| PS | Chloro | 2.539 | 1.504 | 132.914 | 2.229 | 0.0012 |
| PS | Chloro | 2.017 | 2.287 | 133.814 | 2.26 | 0.0013 |
| PS | Chloro | 0.981 | 1.124 | 117.069 | 2.083 | 0.0011 |
| PS | Chloro | 1.671 | 1.325 | 127.64 | 2.147 | 0.0011 |
| PS | Amino | 2.623 | 1.058 | 132.384 | 2.143 | 0.0014 |
| PS | Amino | 2.854 | 1.904 | 133.492 | 2.22 | 0.0018 |
| PS | Amino | 2.57 | 1.777 | 132.922 | 2.215 | 0.0018 |
| PS | Amino | 2.429 | 2.425 | 135.049 | 2.262 | 0.0020 |
| PS | Amino | 2.105 | 2.477 | 134.355 | 2.259 | 0.0020 |
| PS | Amino | 1.633 | 1.356 | 127.898 | 2.125 | 0.0014 |

Once the optimal design is identified, the corresponding processing condition can be obtained by mapping the optimized design variables to processing energy using the PS relationship established in our previous work [40]:

$$\bar{I}_{filler} = f(\text{matrix}) \sinh^2(2\,W_{PF}/W_{FF} - 1)\log(E_\gamma + 1) + C_0, \quad (16)$$

where $\bar{I}_{filler}$ is the normalized interphase area, $f(\text{matrix})$ and $C_0$ are polymer (aka matrix) dependent constants, $W_{PF}/W_{FF}$ is the filler-matrix compatibility descriptors and $E_\gamma$ is the processing energy descriptor that we seek. For illustration, we choose the design (d) in Fig. 10, favoring high breakdown strength of 135.05 kV/mm, as our optimal solution. Microstructure reconstruction corresponding to $VF = 2.43\%$, $\theta = 2.43$ is performed and $\bar{I}_{filler}$ is found to be 0.185. For PS, $f(\text{matrix})$ and $C_0$ are 0.00995 and 0.08798 respectively. For amino-modified Silica nanoparticles dispersed in PS, $W_{PF}/W_{FF} = 0.95$. Plugging these values in Eq. 16 leads to $E_\gamma = 10.47\ kJ/g$. Thus, we can identify designs satisfying application specific material properties and deduce processing parameter necessary for manufacturing.

## 4. CONCLUSIONS

In this article, we presented a data-centric Bayesian Optimization framework for materials design and innovation. Our framework pivots around LVGP, a novel machine learning method that projects qualitative variables onto a continuous latent space for covariance formulation. Uncertainty quantification by LVGP helps in navigating a complex design space using expected improvement acquisition function that balances exploration and exploitation. Visualization of latent space estimated by LVGP provides insight into behavior of levels for each qualitative variable. Generalization of LVGP based BO for multi-objective problems is developed using the expected minimax improvement acquisition criterion. The efficacy of the proposed framework is demonstrated using a case study centered on designing novel insulating nanocomposites with optimal choice of polymer, surface modification and morphology. We considered silica nanoparticles with Amino, Chloro and Octyl surface modifications dispersed in PS, PMMA polymers. Our implementation integrates empirical data with state-of-the-art techniques in interphase modelling, SDF based MCR for dimensionality reduction, and FEA-based structure-property simulations. Design formulation for single and multi-objective Bayesian optimization was presented using two qualitative (polymer and surface modification) and two quantitative (filler volume fraction and dispersion) variables. Both formulations led to similar optimal designs comprising Amino modified silica nanoparticles dispersed in PS matrix with low volume fraction and dispersion. The relative positioning of surface modification levels in latent space corroborates existing knowledge about their behavior. Processing energy required for fabrication of optimal design was evaluated using processing to structure mapping, to complete the bi-directional traversal across PSP paradigms and demonstrate the material genome approach to material design. While LVGP based BO and MOBO are applicable for any engineering design problem, their ability to facilitate concurrent optimization of composition and microstructure w.r.t. one or more properties, makes them a powerful tool for design of functional materials.

For future work, developing accurate simulation models based on Molecular Dynamics and Density Functional Theory is necessary for understanding & evaluating material properties such as Dielectric breakdown strength and interphase behavior. Additionally, we are working on continuously expanding the capabilities of the NanoMine that allows the exploration of curated data gathered from literature and the exploitation of data in the proposed BO framework.

## ACKNOWLEDGEMENT
Support from NSF grants (ACI 1640840, CMMI 1729452, CMMI 1818574, CMMI 1729743, CMMI 1537641, OAC 1835782) and Center for Hierarchical Materials Design (ChiMaD NIST 70NANB14H012) are greatly appreciated.